\documentclass[11pt]{iopart}
\expandafter\let\csname equation*\endcsname\relax
\expandafter\let\csname endequation*\endcsname\relax
\usepackage{times,amsmath,amssymb,graphicx,cite,multirow,color,url,subfigure,booktabs,makecell,afterpage,threeparttable}
\usepackage[lined,ruled,boxed]{algorithm2e}

\usepackage{CJKutf8}
\usepackage[lined,ruled]{algorithm2e}
\allowdisplaybreaks

\makeatletter
\def\@makefnmark{\hbox{\@textsuperscript{\normalfont\@thefnmark}}}
\renewcommand\@makefntext[1]{%
  \noindent
  \hb@xt@1em{\hss\@textsuperscript{\normalfont\@thefnmark}} #1}

\makeatother
\begin{document}
\title{Magnetoencephalography (MEG) Based Non-Invasive Chinese Speech Decoding}

\author{Zhihong~Jia$^{1}$, Hongbin~Wang$^{1}$, Yuanzhong Shen$^{2}$, Feng Hu$^{2}$, Jiayu~An$^{1}$, Kai Shu$^{2,*}$ and Dongrui~Wu$^{1,*}$}
\address{$^1$ Ministry of Education Key Laboratory of Image Processing and Intelligent Control, School of Artificial Intelligence and Automation, Huazhong University of Science and Technology, Wuhan 430074, China.}
\address{$^2$ Department of Neurosurgery, Tongji Hospital, Tongji Medical College, Huazhong University of Science and Technology, Wuhan 430030, China.}
\address{$^*$ Author to whom any correspondence should be addressed.}
\ead{kshu@tjh.tjmu.edu.cn, drwu09@gmail.com.}
\vspace{10pt}

\begin{abstract}
As an emerging paradigm of brain-computer interfaces (BCIs), speech BCI has the potential to directly reflect auditory perception and thoughts, offering a promising communication alternative for patients with aphasia. Chinese is one of the most widely spoken languages in the world, whereas there is very limited research on speech BCIs for Chinese language. This paper reports a text-magnetoencephalography (MEG) dataset for non-invasive Chinese speech BCIs. It also proposes a multi-modality assisted speech decoding (MASD) algorithm to capture both text and acoustic information embedded in brain signals during speech activities. Experiment results demonstrated the effectiveness of both our text-MEG dataset and our proposed MASD algorithm. To our knowledge, this is the first study on modality-assisted decoding for non-invasive speech BCIs.
\end{abstract}

\vspace{2pc}

\section{Introduction}

A brain-computer interface (BCI) establishes a direct communication pathway between the human brain and external devices, with the goal of reflecting the user's intentions \cite{Mridha2021}. It is widely used in neuroscience and offers the potential to help paralysed patients restore motor abilities and improve communication abilities.

There are multiple commonly used BCI paradigms, e.g., steady-state visually evoked potential (SSVEP), event-related potential (ERP), motor imagery (MI), etc. \cite{Abiri2019}. However, SSVEP and ERP are prone to visual fatigue, and users with visual impairment are unable to use them. The MI paradigm requires subject training and is limited to carrying out simple directives. Furthermore, some users suffer from ``BCI illiteracy'' \cite{Dickhaus2009}. Therefore, a more comfortable and efficient BCI paradigm is still in need.

\begin{figure}[htbp]\centering
\includegraphics[width=.6\linewidth,clip]{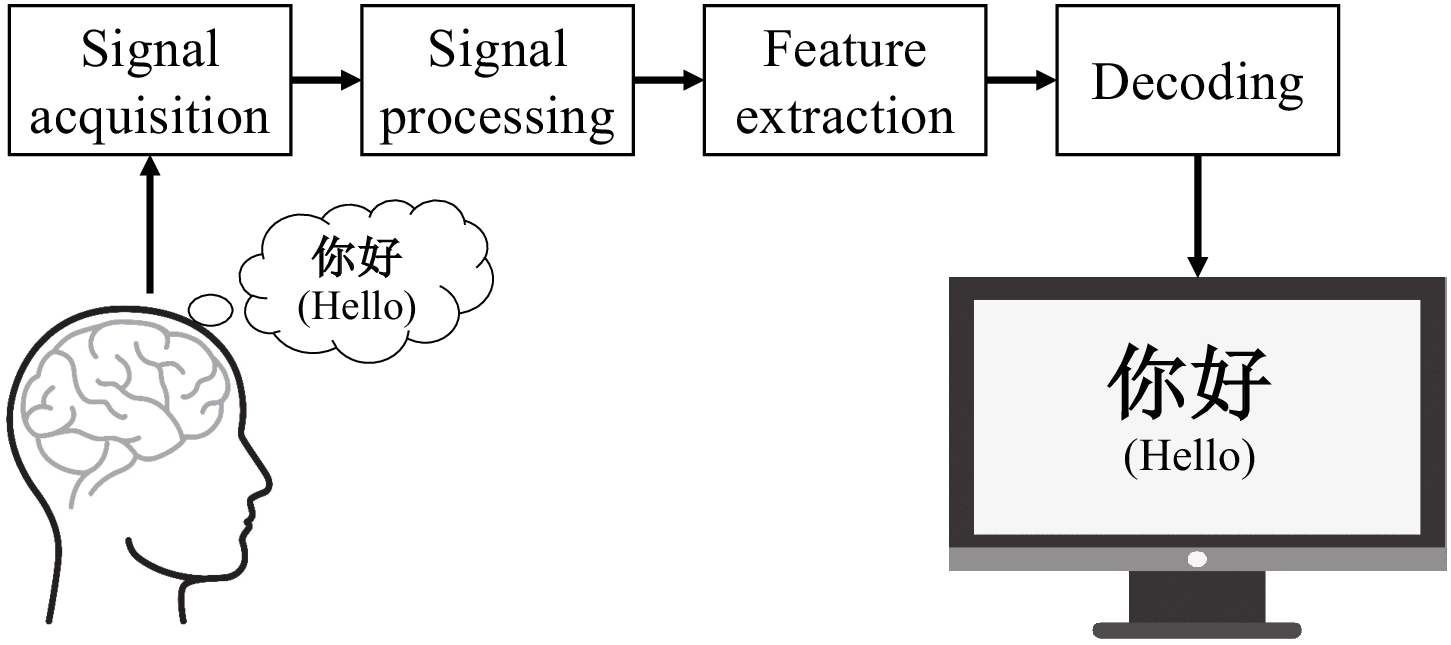}
\caption{Flowchart of a speech BCI system.} \label{fig:system}
\end{figure}

Speech BCI, as an emerging paradigm of BCIs, uses algorithms to translate brain activities into communication signals directly \cite{Brumberg2010}. These signals can take the form of text (e.g., words or sentences displayed on a screen), acoustics (e.g., vocalized sounds or phrases), or facial movements that accompany speech \cite{Silva2024}. This paradigm does not require subject training or prolonged screen gaze, allowing for an immediate reflection of what the user hears and/or thinks. It requires no stimulation, which is more user-friendly. Compared with traditional paradigms, it enables much more commands, allowing more effective communication and device control. Its emergence offers a ``speech prosthetic'' to those with speech impairments due to neurological injuries, giving hope to the severely affected \cite{Anumanchipalli2019}. Fig.~\ref{fig:system} shows the flowchart of a speech BCI system.

As one of the most widely spoken languages in the world, Chinese has over 1.3 billion native speakers \cite{Yaqoub2023}. However, there are very few studies that focus on Chinese speech BCI. Existing research tends to concentrate on the phoneme level and small vocabularies, making it difficult to achieve even basic communication. In addition, most current research relies on invasive BCIs, which require surgical implantation, increasing the risk and cost.

Magnetoencephalography (MEG) measures magnetic fields on the surface of the skull non-invasively \cite{Wang2023}. Compared to electroencephalography (EEG), MEG offers more stable signal quality, as well as superior temporal and spatial resolution \cite{Veena2020}. Therefore, we acquire a non-invasive MEG based Chinese speech BCI dataset in this paper.

Traditional machine learning methods typically require handcrafted feature extraction, which is time-consuming and relies on expert knowledge and experience \cite{DaSalla2009}. End-to-end deep neural networks can automatically extract features directly from raw brain signals, simplifying feature extraction \cite{Zhang2024}. A commonly used speech BCI decoding approach is to focus on the language generation process, decoding from the most fundamental acoustic units, and applying matching rules to form sentences. The sentences are then corrected through a language model to obtain the final results \cite{Willett2021, Willett2023, Metzger2023}. Another feasible decoding approach employs modality assistance, using features from other modalities to guide brain feature extractor training \cite{Defossez2023}. This approach can capture brain features more efficiently and comprehensively, without relying too much on prior knowledge.

This paper focuses on non-invasive Chinese speech BCI decoding. We establish an MEG based Chinese speech BCI dataset, and propose a multi-modality assisted speech decoding (MASD) algorithm. MASD integrates both text and synthetic speech features to constrain the consistency of modal feature representations, enhancing the MEG feature extractor. Our main contributions are:
\begin{enumerate}
\item We designed a Chinese speech BCI experiment with a 48-word corpus, and acquired a text-MEG dataset from nine subjects.
\item We proposed an MASD algorithm that utilizes different modalities to assist MEG feature extraction, which can capture brain features effectively with limited training data and significantly improve the decoding performance.
\item We investigated effective data augmentation approaches in speech BCI.
\end{enumerate}

The remainder of this paper is organized as follows: Section~\ref{sect:Related Work} introduces related works in speech BCIs. Section~\ref{sect:Material} describes our dataset. Section~\ref{sect:Method} introduces our MASD algorithm. Section~\ref{sect:Experiment} presents the experiment results. Finally, Section~\ref{sect:Conclusions} draws conclusions.

\section{Related Works} \label{sect:Related Work}

This section briefly reviews previous research on speech BCIs.

\subsection{Speech BCI Experiments}

According to the literature \cite{Seikel2023}, the auditory language center (part of Wernicke's area) is located in the posterior superior temporal gyrus, the motor language center (Broca's area) is located in the posterior inferior frontal gyrus, the reading center (which includes a portion of Wernicke's area and the angular gyrus above it) is located in the angular gyrus of the inferior parietal lobe, and the writing language center is located in the middle frontal gyrus. However, the brain's speech areas are not completely clear. Willett et al. \cite{Willett2023} pointed out that Broca's area contains minimal information related to orofacial movements, phonemes, or words, and lacks neural activity associated with speech production. Venezia et al. \cite{Venezia2019} noted that aphasia is usually caused by stroke or neurodegeneration in cortical areas such as the superior temporal gyrus, the supramarginal gyrus, and the precentral gyrus. Bouchard et al. \cite{Bouchard2013} indicated that neurons in the ventral sensorimotor cortex and the central precentral gyrus control vocal tract movements, coordinated with exhalation to generate sound waves.

There have been several experiments on speech BCIs, mostly on invasive BCIs in English language.

Willett et al. \cite{Willett2021} implanted four microelectrode arrays on the brain surface of paralyzed patients, and recorded neural activities during imagined handwriting tasks to achieve real-time decoding. Tankus et al. \cite{Tankus2012} investigated the neural encoding mechanisms of language. They employed deep electrodes to record intracranial EEG signals as subjects spoke specific vowels or simple syllables aloud. Moses et al. \cite{Moses2021} constructed a corpus of 50 words and 50 sentences, capturing electrocorticogram (ECoG) signals during reading aloud tasks. Metzger et al. \cite{Metzger2023} designed speech and non-speech tasks, and collected ECoG signals. The goals were to determine whether a subject was engaged in speech activities, and to decode the speech based on a corpus ranging from 50 to 1024 entries. Verwoert et al. \cite{Verwoert2022} recorded stereo-EEG (sEEG) signals during reading aloud tasks, and synthesized speech from them.

D\'efossez et al. \cite{Defossez2023} used non-invasive MEG and EEG signals recorded during speech perception tasks to identify the corresponding speech segments.

There is very limited research on speech BCIs in Chinese language, and they considered simple syllable decoding. Liu et al. \cite{Liu2023} used ECoG electrodes during awake language mapping to decode an 8-class reading aloud task. The task involved four tones of the syllables ``mi'' and ``ma'', achieving an accuracy of 63.6\%. Ni et al. \cite{Ni2023} recorded EEG signals as subjects listened to the syllables ``ba'', ``yao'' and ``yuan'' with four tones, analyzing Mandarin perception through temporal fine structure and envelope.

\subsection{Speech BCI Decoding}

As one of the earliest works, DaSalla et al. \cite{DaSalla2009} used common spatial patterns as the feature extractor and support vector machine as the classifier, achieving 62.7\% accuracy in 2-vowel classification.

Later studies mostly used deep learning approaches. Jim\'enez-Guarneros et al. \cite{JimenezGuarneros2021} used bidirectional recurrent neural networks (RNNs) in unsupervised domain adaptation, achieving 69.48\% accuracy in three-class classification. Willett et al. \cite{Willett2023} utilized an RNN model for phoneme decoding, employing Viterbi search as the matching rule and a language model for error correction, achieving a 9.1\% word error rate on a 50-word vocabulary and a 23.8\% word error rate on a 125,000-word vocabulary. Metzger et al. \cite{Metzger2023} utilized a bidirectional RNN model for phoneme decoding, employing connectionist temporal classification beam search as the matching rule and a language model for error correction, achieving a median word error rate of 25\%.

Research on modality-assisted approaches in speech BCI is very limited. D\'efossez et al. \cite{Defossez2023} used contrastive learning to match invasive brain signals and audio segments, achieving 41\% accuracy across more than 1,000 candidate segments. This paper is the first to study modality-assisted decoding for non-invasive speech BCIs.

\subsection{Modality-Assisted Decoding in Other BCI Paradigms}

Modality-assisted decoding is more commonly used in other BCI paradigms to enhance the decoding performance. Benchetrit et al. \cite{Benchetrit2024} aligned MEG data with visual embeddings from a pretrained model, achieving a top-5 accuracy of 70.33\% in 200-class classification. Song et al. \cite{Song2024} aligned EEG data with image features, achieving a top-1 accuracy of 15.6\% in a 200-class zero-shot task. Du et al. \cite{Du2023} utilized multimodal signals within the Brain-Visual-Linguistic framework to train the brain feature extractor, demonstrating that models integrating visual and linguistic features outperformed single-feature models.

\section{Dataset}   \label{sect:Material}

This section describes our data acquisition process. The dataset is available upon request.

This study was approved on 4/30/2024 by the Medical Ethics Committee of Tongji Hospital, affiliated to Tongji Medical College of Huazhong University of Science and Technology (Ethics Approval Number: TJ-IRB202404079). It was conducted in accordance with the principles embodied in the Declaration of Helsinki and in accordance with local statutory requirements. All participants gave written informed consent to participate in the study.

Brain signals were recorded using a helium-free MEG system manufactured by Beijing X-Magtech Technologies.

\subsection{Subjects}

All nine subjects (7m/2f, 12-60 years old) were native Mandarin speakers and right-handed, had completed junior high school education, with normal or corrected-to-normal eyesight (1.0 or higher) and normal hearing/reading/cognitive abilities. They were free of claustrophobia or any other conditions that could pose risks during data collection.

\subsection{Task Design}

Our corpus includes 48 high-frequency Chinese words, covering everyday vocabulary and common medical terms, as shown in Table~\ref{table:corpus}. They were selected by considering the usage frequencies of common Chinese words in the Inter-language Corpus of Chinese from Global Learners \cite{Zhang2013} and the Beijing Language and Culture University Corpus \cite{Xun2016}, and inspired by previous research \cite{Moses2021, Verwoert2022, Willett2023}. The corpus comprehensively covers all phonemes in Mandarin, including all types of initials, finals, and tones. Particularly, the initials and tones are balanced.

\begin{table}[htbp]     \centering  \setlength{\tabcolsep}{0.8mm}
\caption{The 48 Chinese words used in our experiments.}    \label{table:corpus}
\scalebox{0.68}{
\begin{tabular}{@{\hspace{0.5em}}cccccc@{\hspace{0.5em}}}
\toprule
Word & Tone & Initial & Initial\_8 & Final & Final class \\ \midrule
\begin{CJK}{UTF8}{gbsn}飘\end{CJK}     & T1   & p       & I1         & iao   & F2          \\
\begin{CJK}{UTF8}{gbsn}方\end{CJK}     & T1   & f       & I2         & ang   & F1          \\
\begin{CJK}{UTF8}{gbsn}丢\end{CJK}     & T1   & d       & I4         & iu    & F2          \\
\begin{CJK}{UTF8}{gbsn}他\end{CJK}     & T1   & t       & I4         & a     & F1          \\
\begin{CJK}{UTF8}{gbsn}天\end{CJK}     & T1   & t       & I4         & ian   & F2          \\
\begin{CJK}{UTF8}{gbsn}光\end{CJK}     & T1   & g       & I7         & uang  & F3          \\
\begin{CJK}{UTF8}{gbsn}均\end{CJK}     & T1   & j       & I6         & vn    & F4          \\
\begin{CJK}{UTF8}{gbsn}缺\end{CJK}     & T1   & q       & I6         & ve    & F4          \\
\begin{CJK}{UTF8}{gbsn}出\end{CJK}     & T1   & ch      & I5         & u     & F3          \\
\begin{CJK}{UTF8}{gbsn}说\end{CJK}     & T1   & sh      & I5         & uo    & F3          \\
\begin{CJK}{UTF8}{gbsn}扔\end{CJK}     & T1   & r       & I5         & eng   & F1          \\
\begin{CJK}{UTF8}{gbsn}三\end{CJK}     & T1   & s       & I3         & an    & F1          \\
\begin{CJK}{UTF8}{gbsn}没\end{CJK}     & T2   & m       & I1         & ei    & F1          \\
\begin{CJK}{UTF8}{gbsn}明\end{CJK}     & T2   & m       & I1         & ing   & F2          \\
\begin{CJK}{UTF8}{gbsn}年\end{CJK}     & T2   & n       & I4         & ian   & F2          \\
\begin{CJK}{UTF8}{gbsn}结\end{CJK}     & T2   & j       & I6         & ie    & F2          \\
\begin{CJK}{UTF8}{gbsn}穷\end{CJK}     & T2   & q       & I6         & iong  & F2          \\
\begin{CJK}{UTF8}{gbsn}床\end{CJK}     & T2   & ch      & I5         & uang  & F3          \\
\begin{CJK}{UTF8}{gbsn}人\end{CJK}     & T2   & r       & I5         & en    & F1          \\
\begin{CJK}{UTF8}{gbsn}昨\end{CJK}     & T2   & z       & I3         & uo    & F3          \\
\begin{CJK}{UTF8}{gbsn}从\end{CJK}     & T2   & c       & I3         & ong   & F1          \\
\begin{CJK}{UTF8}{gbsn}才\end{CJK}     & T2   & c       & I3         & ai    & F1          \\
\begin{CJK}{UTF8}{gbsn}随\end{CJK}     & T2   & s       & I3         & ui    & F3          \\
\begin{CJK}{UTF8}{gbsn}而\end{CJK}     & T2   & -       & I8         & er    & F1          \\
\begin{CJK}{UTF8}{gbsn}把\end{CJK}     & T3   & b       & I1         & a     & F1          \\
\begin{CJK}{UTF8}{gbsn}品\end{CJK}     & T3   & p       & I1         & in    & F2          \\
\begin{CJK}{UTF8}{gbsn}你\end{CJK}     & T3   & n       & I4         & i     & F2          \\
\begin{CJK}{UTF8}{gbsn}旅\end{CJK}     & T3   & l       & I4         & v     & F4          \\
\begin{CJK}{UTF8}{gbsn}两\end{CJK}     & T3   & l       & I4         & iang  & F2          \\
\begin{CJK}{UTF8}{gbsn}缓\end{CJK}     & T3   & h       & I7         & uan   & F3          \\
\begin{CJK}{UTF8}{gbsn}选\end{CJK}     & T3   & x       & I6         & van   & F4          \\
\begin{CJK}{UTF8}{gbsn}准\end{CJK}     & T3   & zh      & I5         & un    & F3          \\
\begin{CJK}{UTF8}{gbsn}早\end{CJK}     & T3   & z       & I3         & ao    & F1          \\
\begin{CJK}{UTF8}{gbsn}有\end{CJK}     & T3   & y       & I8         & ou    & F2          \\
\begin{CJK}{UTF8}{gbsn}我\end{CJK}     & T3   & w       & I8         & o     & F3          \\
\begin{CJK}{UTF8}{gbsn}耳\end{CJK}     & T3   & -       & I8         & er    & F1          \\
\begin{CJK}{UTF8}{gbsn}不\end{CJK}     & T4   & b       & I1         & u     & F3          \\
\begin{CJK}{UTF8}{gbsn}费\end{CJK}     & T4   & f       & I2         & ei    & F1          \\
\begin{CJK}{UTF8}{gbsn}第\end{CJK}     & T4   & d       & I4         & i     & F2          \\
\begin{CJK}{UTF8}{gbsn}个\end{CJK}     & T4   & g       & I7         & e     & F1          \\
\begin{CJK}{UTF8}{gbsn}看\end{CJK}     & T4   & k       & I7         & an    & F1          \\
\begin{CJK}{UTF8}{gbsn}快\end{CJK}     & T4   & k       & I7         & uai   & F3          \\
\begin{CJK}{UTF8}{gbsn}话\end{CJK}     & T4   & h       & I7         & ua    & F3          \\
\begin{CJK}{UTF8}{gbsn}下\end{CJK}     & T4   & x       & I6         & ia    & F2          \\
\begin{CJK}{UTF8}{gbsn}这\end{CJK}     & T4   & zh      & I5         & e     & F1          \\
\begin{CJK}{UTF8}{gbsn}上\end{CJK}     & T4   & sh      & I5         & ang   & F1          \\
\begin{CJK}{UTF8}{gbsn}要\end{CJK}     & T4   & y       & I8         & ao    & F2          \\
\begin{CJK}{UTF8}{gbsn}问\end{CJK}     & T4   & w       & I8         & en    & F3          \\ \bottomrule
\end{tabular}
}
\begin{tablenotes}
\footnotesize
\item \textasteriskcentered~Initials and finals can be categorized based on the phonetic characteristics of Mandarin.
\begin{enumerate}
\item Based on the place of articulation, initials can be divided into eight classes (Initial\_8): bilabials (I1: /b/, /p/, /m/), labiodentals (I2: /f/), apical anterior affricates and fricatives (I3: /z/, /c/, /s/), alveolars (I4: /d/, /t/, /n/, /l/), retroflexes (I5: /zh/, /ch/, /sh/, /r/), palatals (I6: /j/, /q/, /x/), velars (I7: /g/, /k/, /h/), and others (I8).
\item Based on the structure of oral cavity, finals can be classified into four classes: open finals (F1: not begin with /i/, /u/, or /v/), apical front finals (F2: begin with /i/), back rounded finals (F3: begin with /u/), rounded front finals (F4: begin with /v/).
\end{enumerate}
\end{tablenotes}
\end{table}

\begin{figure*}[htbp]\centering
\includegraphics[width=\linewidth,clip]{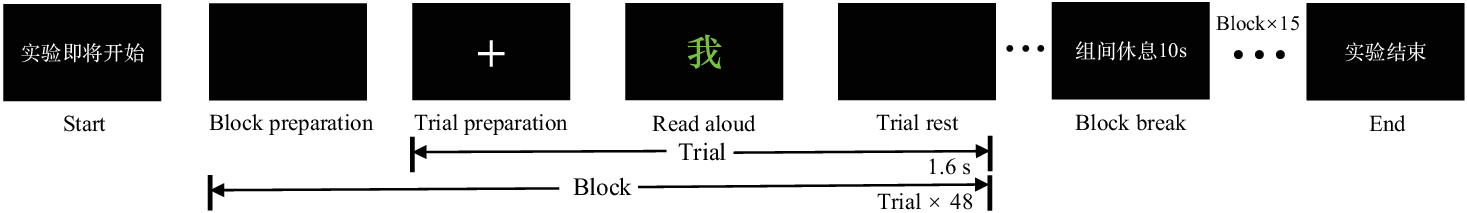}
\caption{Our Chinese speech BCI experiment pipeline.} \label{fig:paradigm}
\end{figure*}

The experiment pipeline, shown in Fig.~\ref{fig:paradigm}, was developed using the open source software package Psychopy \cite{Peirce2019}. We recorded MEG data as the subjects were reading the words on the screen. The experiment began with a button press once the subject was ready. Each word stimulus was preceded by a 0.1 s preparation period. The stimulus was then displayed for 1 s, during which the subject was instructed to read the word aloud. Afterward, there was a 0.5 s rest to account for variations in subjects' reading speeds. Each trial lasted 1.6 s, with all 48 words displayed as a block, followed by a 10 s break. Each subject completed 15 blocks, resulting in a total of 720 trials.

\subsection{Data Processing}

The MEG data were detrended, re-referenced, and then filtered by a 70-170 Hz bandpass filter and a 50 Hz notch filter, corresponding to the typical frequency range for speech BCI decoding \cite{Verwoert2022}. The data were then scaled and clamped before the Hilbert envelope was extracted \cite{Moses2021}. Finally, they were downsampled to 200 Hz.

\section{Method}    \label{sect:Method}

\begin{figure*}[htbp]\centering
\includegraphics[width=\linewidth,clip]{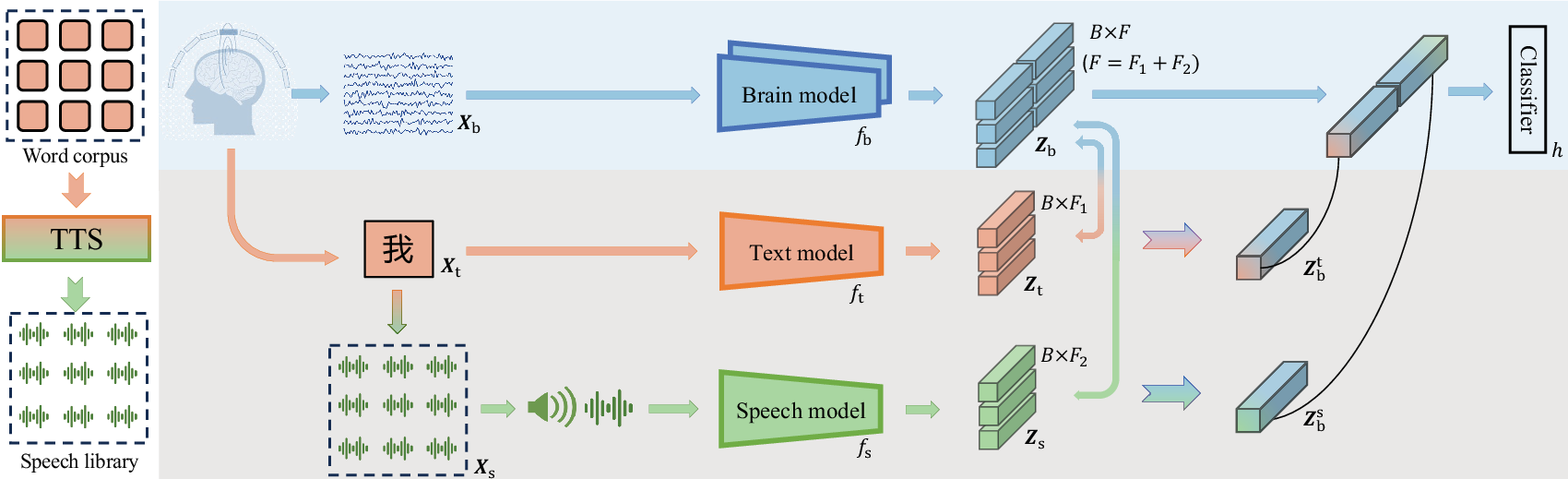}
\caption{Overview of the MASD algorithm.} \label{fig:framework}
\end{figure*}

This section introduces our proposed MASD algorithm, as illustrated in Fig.~\ref{fig:framework}.

\subsection{Overview}

During training, stimulus-response pairs consisting of text as the stimulus and MEG signals as the response are used as inputs. Let $\boldsymbol{X_\text{b}} \in \mathbb{R}^{C \times T}$ be a trial of the MEG signal, where $C$ is the number of channels and $T$ the number of time-domain sampling points. The stimulus text, denoted as $\boldsymbol{X_\text{t}}$, serves as the corresponding label. The MEG model and the text model extract relevant features from $\boldsymbol{X_\text{b}}$ and $\boldsymbol{X_\text{t}}$, respectively. To account for the relationship between speech and brain activities, Text-to-Speech (TTS) technology is employed to generate the corresponding speech signal $\boldsymbol{X_\text{s}}$ from $\boldsymbol{X_\text{t}}$. The speech model extracts the speech features from $\boldsymbol{X_\text{s}}$. The corresponding features of text-brain (speech-brain) are aligned by contrastive learning to optimize the MEG feature extractor.

During inference (test), only the MEG signals are used as input, and they are classified using the classifier head $h$.

\subsection{Speech Library}

To capture acoustic information in brain signals during speech activities, TTS technology is employed to introduce the speech modality $\boldsymbol{X_\text{s}}$ \cite{Kaur2023}. TTS analyzes the input text and constructs a statistical model based on the textual information to predict speech parameters, such as fundamental frequencies and formant frequencies. Subsequently, a vocoder module converts these parameters into a model for speech generation. Baidu's intelligent cloud voice synthesis technology\footnote{\url{https://cloud.baidu.com/product/speech/tts/}} is used to convert the 48-word corpus into a speech library, which is later used to extract speech features.

\subsection{Feature Extraction}

The popular EEGNet \cite{Lawhern2018} is used as the brain model $f_\text{b}$ to extract brain features $\boldsymbol{Z_\text{b}} = f_\text{b}(\boldsymbol{X_\text{b}}) \in \mathbb{R}^{F}$. It is a compact convolutional neural network comprising temporal and spatial convolutions. By employing depthwise and separable convolutions, it significantly reduces the number of parameters.

Specifically, one EEGNet model is used to extract $\boldsymbol{Z_\text{b}^\text{t}} \in \mathbb{R}^{F_1}$, and another is used to extract $\boldsymbol{Z_\text{b}^\text{s}} \in \mathbb{R}^{F_2}$. These two feature representations are concatenated to form the final brain feature representation $\boldsymbol{Z_\text{b}} = [\boldsymbol{Z_\text{b}^\text{t}}, \boldsymbol{Z_\text{b}^\text{s}}] \in \mathbb{R}^{F_1 + F_2}$.

Two pretrained large language models, fastText \cite{Joulin2017} and BERT \cite{Devlin2019}, are used as the text model $f_\text{t}$ to extract text features $\boldsymbol{Z_\text{t}} = f_\text{t}(\boldsymbol{X_\text{t}}) \in \mathbb{R}^{F_1}$:
\begin{enumerate}
\item \emph{fastText} learns word vectors through n-grams. It decomposes words into character-level n-grams and generates word vectors for them to represent the words. This model is particularly beneficial for Chinese language due to its character-based structure.

\item \emph{BERT} stacks multiple transformer encoder layers. Each encoder layer integrates a self-attention mechanism and a feedforward neural network, allowing the model to capture complex dependencies within the input sequence. By combining token/segment/positional embeddings, BERT can comprehensively capture the semantic and contextual information of the text, providing a strong foundational representation for various natural language processing tasks.
\end{enumerate}

Three acoustic feature extractors, including one that computes the traditional Mel spectrogram \cite{Davis1980}, and two pretrained large speech models (wav2vec 2.0 \cite{Baevski2020} and HuBERT \cite{Hsu2021}), are used to extract speech features $\boldsymbol{Z_\text{s}} = f_\text{s}(\boldsymbol{X_\text{s}}) \in \mathbb{R}^{F_2}$:
\begin{enumerate}
\item \emph{Mel spectrogram}. The human auditory system is more sensitive to low-frequency sounds, and converting linear frequencies to the Mel-scale better approximates this characteristic. The conversion between the Mel-scale $m$ and the frequency scale $\nu$ is:
\begin{align}
m=2595\, \log_{10}\left(1+\frac{\nu}{700}\right).
\end{align}

\item \emph{wav2vec 2.0 (wav2vec2)}. It is an end-to-end self-supervised model widely used for speech feature extraction. A convolutional neural network based feature extractor encodes the raw speech waveform into a sequence of frame-level representations, which are subsequently passed to a Transformer to generate contextual representations. wav2vec2 can effectively capture high-quality speech representations.

\item \emph{HuBERT}. HuBERT has a similar architecture as wav2vec2, and adopts a BERT-inspired methodology for self-supervised pretraining.
\end{enumerate}

\subsection{Training Loss}

InfoNCE loss \cite{Oord2018} is used as the contrastive loss. The feature outputs of different modalities are first normalized, and the cosine similarity between text-brain (speech-brain) pairs is then calculated. The distribution is smoothed using temperature scaling. The matching text-brain (speech-brain) pairs are treated as positive pairs, whereas the mismatching pairs serve as negative pairs.

Contrastive learning is employed during iterative training using $\mathcal{L}_\text{t}$ and $\mathcal{L}_\text{s}$, which minimize the distance between the representations of positive pairs while maximizing the distance between the representations of negative pairs:
\begin{align}
\mathcal{L}_\text{t}&=-\frac{1}{B} \sum_{i=1}^{B}\log \frac{\exp \left(\operatorname{sim}\left(\boldsymbol{Z_{\text{b}_i}^\text{t}}, \boldsymbol{Z_{\text{t}_i}^{+}}\right)/\tau\right)}{\sum_{j=1}^{B} \exp \left(\operatorname{sim}\left(\boldsymbol{Z_{\text{b}_i}^\text{t}}, \boldsymbol{Z_{\text{t}_j}}\right)/\tau\right)},  \label{equ: text loss}\\
\mathcal{L}_\text{s}&=-\frac{1}{B} \sum_{i=1}^{B}\log \frac{\exp \left(\operatorname{sim}\left(\boldsymbol{Z_{\text{b}_i}^\text{s}}, \boldsymbol{Z_{\text{s}_i}^{+}}\right)/\tau\right)}{\sum_{j=1}^{B} \exp \left(\operatorname{sim}\left(\boldsymbol{Z_{\text{b}_i}^\text{s}}, \boldsymbol{Z_{\text{s}_j}}\right)/\tau\right)},  \label{equ: speech loss}
\end{align}
where $B$ is the number of input sample pairs in a batch, $\boldsymbol{Z_{\text{b}_i}^\text{t}}$($\boldsymbol{Z_{\text{b}_i}^\text{s}}$) is the latent feature extracted from the current brain sample, $\boldsymbol{Z_{\text{t}_i}^+}$($\boldsymbol{Z_{\text{s}_i}^+}$) contains the text (speech) features associated with the brain data, and $\tau$ is the temperature coefficient.

The overall training loss $\mathcal{L}$ is:
\begin{align}
\mathcal{L}=\mathcal{L}_{\text {CE}}+\lambda_\text{t} \mathcal{L}_\text{t}+\lambda_\text{s} \mathcal{L}_\text{s},    \label{equ: total loss}
\end{align}
where $\mathcal{L}_{\text {CE}}$ is the cross-entropy loss, and $\lambda_\text{t}$ and $\lambda_\text{s}$ are trade-off hyper-parameters.

Algorithm~\ref{alg:MultimodalDecoding} gives the pseudo-code of the MASD algorithm.

\begin{algorithm}[htbp]
\KwIn{$\boldsymbol{X_\text{b}}$: brain signals\;
      \hspace*{12mm} $\boldsymbol{X_\text{t}}$: text signals\;
      \hspace*{12mm} $\text{TTS}$: Text-to-Speech technology\;
      \hspace*{12mm} $f_\text{b}$: brain model\;
      \hspace*{12mm} $f_\text{t}$: text model\;
      \hspace*{12mm} $f_\text{s}$: speech model\;
      \hspace*{12mm} $h$: classifier\;}
\tcp{Speech synthesis}
$\boldsymbol{X_\text{s}} \gets \text{TTS}(\boldsymbol{X_\text{t}})$\;
\tcp{Feature extraction}
$\boldsymbol{Z_\text{b}} \gets f_\text{b}(\boldsymbol{X_\text{b}})$\;
$\boldsymbol{Z_\text{t}} \gets f_\text{t}(\boldsymbol{X_\text{t}})$\;
$\boldsymbol{Z_\text{s}} \gets f_\text{s}(\boldsymbol{X_\text{s}})$\;
\tcp{Classification}
$logits \gets h(\boldsymbol{Z_\text{b}})$\;
\tcp{Calculate loss}
Calculate $\mathcal{L}_\text{t}$ and $\mathcal{L}_\text{s}$ by (\ref{equ: text loss}) and (\ref{equ: speech loss}), respectively\;
Calculate $\mathcal{L}$ by (\ref{equ: total loss})\;
\tcp{Parameter Update}
Optimize the parameters of $f_\text{b}$ and $h$ using back-propogation\;
\Return $f_\text{b}$ and $h$\
\caption{The MASD algorithm for speech BCI decoding.} \label{alg:MultimodalDecoding}
\end{algorithm}

\subsection{Data Augmentation}

Noise-based data augmentation is used to expand the training dataset and increase the classification accuracy.

For noise augmentation in the time domain, noise is directly added to the Hilbert envelope extracted from the MEG signal. For noise augmentation in the frequency domain, we perform fast Fourier transform to the envelope, add noise to the coefficients, and then perform inverse Fourier transform to generate the augmented MEG data in the time domain.

Let $\xi$ be the MEG data before augmentation. The following four types of noise are considered:
\begin{enumerate}
\item \emph{Gaussian Noise} \cite{Boyat2015}, also known as white noise, which simulates random interference. Its probability density function follows a Gaussian distribution:
\begin{align}
p(\xi)=\frac{1}{\sqrt{2 \pi} \sigma} \exp \left(-\frac{(\xi-\mu)^{2}}{2 \sigma^{2}}\right),
\end{align}
where $\mu$ and $\sigma$ denote the mean and standard deviation of the noise, respectively.

\item \emph{Poisson Noise} \cite{Boyat2015}, which simulates nonlinear interferences or occasional events. It is characterized by a Poisson distribution:
\begin{align}
p(\xi)=\frac{\kappa^{\xi} e^{-\kappa}}{\xi!},
\end{align}
where $\kappa$ is the rate or intensity of events occurring in a fixed time interval.

\item \emph{Pink Noise} \cite{Boyat2015}, which is frequently employed to replicate intricate noise patterns observed in natural and biological systems. It is characterized by a spectral density inversely proportional to the frequency $\nu$, i.e., lower frequencies contain higher energy, whereas higher frequencies possess comparatively lower energy:
\begin{align}
p(\nu) \propto \frac{1}{\nu^{\alpha}},
\end{align}
where $\alpha$ is an adjustable parameter.

\item \emph{Salt \& Pepper Noise} \cite{Boyat2015}, which is commonly used to simulate abrupt and high-amplitude discrete interferences. It is introduced by randomly selecting a small proportion of data points and setting their values to the maximum or minimum:
\begin{align}
\setlength{\arraycolsep}{0.5pt}
\xi_{i}=\left\{\begin{array}{ll}
\max (\xi), & \text{ with probability } p_{\text{s}} \\
\min (\xi), & \text{ with probability } p_{\text{p}} \\
\xi_{i}, & \text{ with probability } 1-p_{\text{s}}-p_{\text{p}}
\end{array}\right.,
\end{align}
where $p_{\text{s}}$ and $p_{\text{p}}$ represent the probabilities of salt and pepper noise, respectively.
\end{enumerate}

\section{Experiments}    \label{sect:Experiment}

This section presents the experiment results to validate the effectiveness of our proposed MASD algorithm. The code is available at \url{https://github.com/ZhihongJia/MASD/tree/main}.

\subsection{Implementation Details}

Five-fold cross-validation was used in within-subject experiments to divide the data into training and testing subsets. The training subset was further partitioned to establish a validation set, resulting in a data distribution ratio of 10:2:3 for training, validation, and testing, respectively.

Leave-one-subject-out cross-validation was used in cross-subject experiments. One-eighth data from each subject in the training subset were randomly selected as the validation set. The resulting ratio of training, validation, and testing data was 7:1:1.

The batch size was 48 in within-subject experiments, and 720 in cross-subject experiments. The temperature coefficient $\tau = 0.01$, and the training process employed AdamW optimizer with learning rate 0.001. The maximum number of training epochs was 100, and early stopping was applied with a patience of 50.

We used accuracy (ACC) as the performance evaluation metric. Due to class-imbalance of the phoneme, we also computed the balanced classification accuracy (BCA: the average of per-class accuracies). All experiments were repeated 20 times, and the average results are reported.

\subsection{Visualizations}

\begin{figure*}[htbp]\centering
\includegraphics[width=\linewidth,clip]{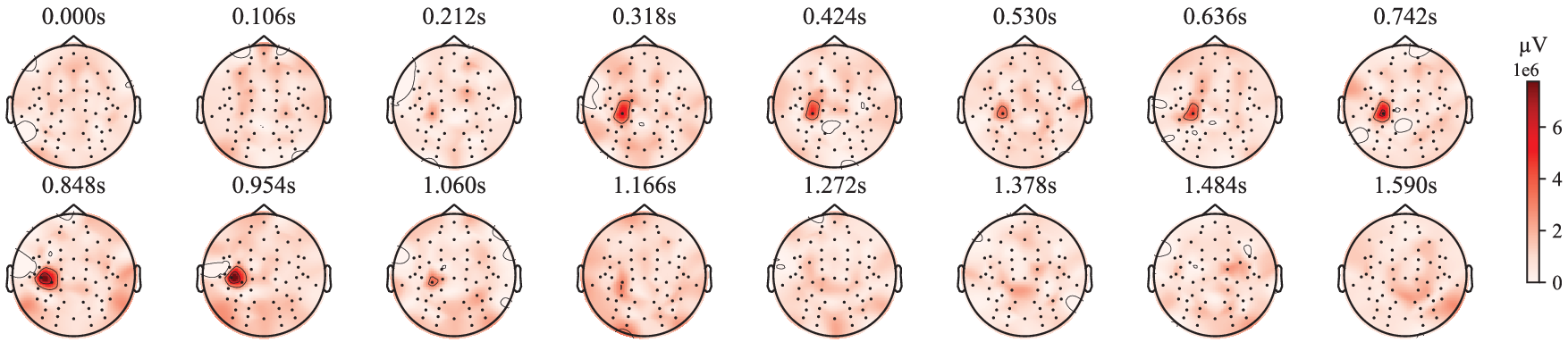}
\caption{Change in brain energy over time during speech activity.} \label{fig:topomap_time}
\end{figure*}

Fig.~\ref{fig:topomap_time} visualizes the change in brain energy over time for a single trial. After 0.1 s of stimulus onset, the corresponding brain regions gradually became active, peaking around 0.9 s. Then, the activities gradually returned to the resting state around 1.6 s. This indicates that our trial length is enough to capture the speech activities.

\begin{figure}[htbp]\centering
\includegraphics[width=0.6\linewidth,clip]{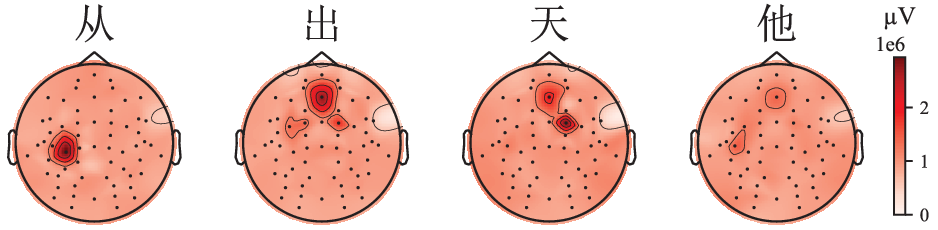}
\caption{Brain region activations for different words.} \label{fig:topomap_word}
\end{figure}

MEG signals were averaged during the entire trial to examine whether the activation of brain regions varies among different words. Fig.~\ref{fig:topomap_word} presents several representative topographical maps. Different words activate different brain regions, consistent with the findings in \cite{Huth2016}. This result suggests the possibility to decode different words from MEG signals, and also the necessity to use channels that cover different regions of the brain.

\subsection{Main Results}

Table~\ref{table:Within-subject} presents the top-5 and top-1 accuracies (\%) in within-subject decoding. The best performance in each column is marked in bold. The random top-1 accuracy for 48-word classification is 2.08\%, and the random top-5 accuracy is 10.40\%. The top-1 and top-5 single-modality decoding accuracies, which used EEGNet to decode the MEG signals only without any modality assistance, were 2.5-3 times higher than random guess, demonstrating that MEG data indeed contain useful speech related information.

\begin{table*}[htbp]     \centering    \setlength{\tabcolsep}{0.7mm}
\caption{Top-5 and top-1 accuracies (\%) in within-subject speech decoding.}  \label{table:Within-subject}
\renewcommand{\arraystretch}{1}
\scalebox{0.59}{
\begin{tabular}{@{}cl|cccccccccccccccccc|cc@{}}
\toprule
\multicolumn{2}{c|}{\multirow{3}{*}[-1.5ex]{Approach}} & \multicolumn{18}{c|}{Subject} & \multicolumn{2}{c}{\multirow{2}{*}[-0.5ex]{Average}} \\ \cmidrule(lr){3-20}
\multicolumn{2}{c|}{} & \multicolumn{2}{c}{1} & \multicolumn{2}{c}{2} & \multicolumn{2}{c}{3} & \multicolumn{2}{c}{4} & \multicolumn{2}{c}{5} & \multicolumn{2}{c}{6} & \multicolumn{2}{c}{7} & \multicolumn{2}{c}{8} & \multicolumn{2}{c|}{9} & \multicolumn{2}{c}{} \\ \cmidrule(l){3-22}
\multicolumn{2}{c|}{} & Top5 & Top1 & Top5 & Top1 & Top5 & Top1 & Top5 & Top1 & Top5 & Top1 & Top5 & Top1 & Top5 & Top1 & Top5 & Top1 & Top5 & Top1 & Top5 & Top1 \\ \midrule
\multicolumn{2}{c|}{Single-modality} & 29.92 & 7.38 & 29.92 & 8.03 & 27.61 & 7.32 & 19.83 & 4.56 & 28.21 & 7.08 & 18.84 & 4.24 & 17.40 & 4.06 & 28.40 & 8.29 & 32.66 & 8.91 & 25.86 & 6.65 \\ \midrule
\multicolumn{1}{c|}{\multirow{6}{*}{MASD}} & fastText+Mel & 36.20 & 9.81 & 37.84 & 11.63 & 34.53 & 10.33 & 26.51 & 6.74 & 35.78 & 10.20 & 22.22 & 5.02 & 24.07 & 5.99 & 36.68 & 12.24 & 41.67 & 12.85 & 32.83 & 9.42 \\
\multicolumn{1}{c|}{} & fastText+wav2vec2 & 36.66 & 9.91 & 39.50 & 11.58 & 34.51 & 10.38 & 26.86 & 6.63 & 35.97 & 10.08 & \textbf{22.68} & 5.18 & \textbf{24.40} & 6.04 & \textbf{38.50} & 12.33 & 43.76 & 14.02 & 33.65 & 9.57 \\
\multicolumn{1}{c|}{} & fastText+HuBERT & 36.67 & 9.88 & 39.22 & \textbf{12.41} & \textbf{35.29} & \textbf{10.71} & 26.72 & \textbf{7.04} & 36.09 & 10.17 & 22.11 & 5.26 & 24.33 & \textbf{6.17} & 38.15 & 12.53 & 44.17 & 13.69 & 33.64 & 9.76 \\
\multicolumn{1}{c|}{} & BERT+Mel & 36.21 & 9.70 & 37.46 & 11.16 & 33.76 & 10.09 & 26.57 & 6.21 & 34.50 & 9.65 & 21.87 & 5.30 & 23.78 & 5.94 & 36.46 & 12.20 & 42.21 & 13.03 & 32.53 & 9.25 \\
\multicolumn{1}{c|}{} & BERT+wav2vec2 & \textbf{37.08} & \textbf{10.03} & \textbf{39.53} & 12.29 & 34.72 & 10.67 & \textbf{27.01} & 6.86 & \textbf{36.86} & \textbf{10.63} & 22.41 & 5.26 & 24.27 & 6.12 & 38.21 & 13.00 & 44.22 & \textbf{14.56} & \textbf{33.81} & \textbf{9.94} \\
\multicolumn{1}{c|}{} & BERT+HuBERT & 37.02 & 9.74 & 39.22 & 12.13 & 34.76 & 10.49 & 26.24 & 6.89 & 36.08 & 10.53 & 22.39 & \textbf{5.36} & 23.65 & 5.86 & 38.13 & \textbf{13.34} & \textbf{44.55} & 14.06 & 33.56 & 9.82 \\ \bottomrule
\end{tabular}}
\end{table*}

Various configurations of our proposed MASD approach consistently improved the decoding accuracy. The average top-5 accuracy was increased by 7.95\%, corresponding to a 7.95/25.86=30.74\% relative improvement. The average top-1 accuracy was improved by 3.29\%, corresponding to a 3.29/6.65=49.47\% relative improvement.

Table~\ref{table:Cross-subject} presents the cross-subject speech decoding results. Similar observations as in within-subject decoding can be made, though the performance improvements were smaller, due to more challenging nature of cross-subject speech decoding.

\begin{table*}[htbp]    \centering     \setlength{\tabcolsep}{0.7mm}
\caption{Top-5 and top-1 accuracies (\%) in cross-subject speech decoding.}   \label{table:Cross-subject}
\renewcommand{\arraystretch}{1}
\scalebox{0.59}{
\begin{tabular}{@{}cl|cccccccccccccccccc|cc@{}}
\toprule
\multicolumn{2}{c|}{\multirow{3}{*}[-1.5ex]{Approach}} & \multicolumn{18}{c|}{Subject} & \multicolumn{2}{c}{\multirow{2}{*}[-0.5ex]{Average}} \\ \cmidrule(lr){3-20}
\multicolumn{2}{c|}{} & \multicolumn{2}{c}{1} & \multicolumn{2}{c}{2} & \multicolumn{2}{c}{3} & \multicolumn{2}{c}{4} & \multicolumn{2}{c}{5} & \multicolumn{2}{c}{6} & \multicolumn{2}{c}{7} & \multicolumn{2}{c}{8} & \multicolumn{2}{c|}{9} & \multicolumn{2}{c}{} \\ \cmidrule(l){3-22}
\multicolumn{2}{c|}{} & Top5 & Top1 & Top5 & Top1 & Top5 & Top1 & Top5 & Top1 & Top5 & Top1 & Top5 & Top1 & Top5 & Top1 & Top5 & Top1 & Top5 & Top1 & Top5 & Top1 \\ \midrule
\multicolumn{2}{c|}{Single-modality} & 16.53 & 3.75 & 14.37 & 2.75 & 15.31 & 3.29 & 11.63 & 2.40 & 28.08 & 6.99 & 22.59 & 4.87 & 17.58 & 3.53 & 19.26 & 4.18 & 25.35 & 5.61 & 18.97 & 4.15 \\ \midrule
\multicolumn{1}{c|}{\multirow{6}{*}{MASD}} & fastText+Mel & \textbf{18.01} & \textbf{4.27} & \textbf{16.10} & 3.47 & 16.01 & 3.51 & \textbf{12.74} & 2.60 & 30.18 & 7.98 & 23.14 & \textbf{5.42} & 18.76 & 3.92 & 19.96 & 4.98 & 28.40 & 6.63 & \textbf{20.37} & 4.75 \\
\multicolumn{1}{c|}{} & fastText+wav2vec2 & 17.14 & 3.70 & 15.39 & 3.67 & 16.26 & \textbf{3.67} & 12.73 & 2.38 & 29.92 & 7.97 & \textbf{23.28} & 5.31 & \textbf{19.57} & \textbf{4.18} & \textbf{20.13} & 5.18 & 28.12 & \textbf{6.83} & 20.28 & \textbf{4.76} \\
\multicolumn{1}{c|}{} & fastText+HuBERT & 17.24 & 4.09 & 15.69 & \textbf{3.77} & 16.05 & 3.45 & 12.42 & 2.40 & \textbf{30.22} & \textbf{8.23} & 23.01 & 5.15 & 19.18 & 3.66 & 20.10 & 5.06 & 28.52 & 6.38 & 20.27 & 4.69 \\
\multicolumn{1}{c|}{} & BERT+Mel & 16.78 & 3.85 & 14.65 & 3.08 & 16.62 & 3.52 & 12.21 & 2.58 & 29.51 & 7.63 & 23.13 & 5.13 & 18.96 & 3.92 & 19.78 & 4.94 & 27.44 & 6.30 & 19.90 & 4.55 \\
\multicolumn{1}{c|}{} & BERT+wav2vec2 & 16.72 & 3.99 & 15.39 & 3.57 & \textbf{16.92} & 3.57 & 12.27 & 2.48 & 29.58 & 8.15 & 23.06 & 5.03 & 19.30 & 3.90 & 19.83 & 5.13 & \textbf{28.60} & 6.26 & 20.18 & 4.67 \\
\multicolumn{1}{c|}{} & BERT+HuBERT & 16.85 & 3.79 & 15.09 & 3.26 & 16.81 & 3.61 & 12.32 & \textbf{2.62} & 29.62 & 8.11 & 23.10 & 5.19 & 19.47 & 3.91 & 19.94 & \textbf{5.27} & 28.16 & 6.33 & 20.15 & 4.68 \\ \bottomrule
\end{tabular}}
\end{table*}

\subsection{Phoneme Decoding Results} \label{subsec: Phoneme}

Considering the inherent properties of Chinese phonetics, we conducted an analysis of the most fundamental acoustic units. Table~\ref{table:Within-subject Phoneme} presents the phoneme decoding results. Employing speech features from synthetic speech improved the classification accuracy in all phoneme decoding experiments, demonstrating again the effectiveness of MASD.

\begin{table*}[htbp]   \centering   \setlength{\tabcolsep}{3.5mm}
\caption{Accuracy (\%) in within-subject phoneme decoding.}    \label{table:Within-subject Phoneme}
\renewcommand{\arraystretch}{1}
\scalebox{0.59}{
\begin{tabular}{@{\hspace{1em}}c|c|ccccccccc|c@{\hspace{1em}}}
\toprule
\multirow{2}{*}[-0.5ex]{Phoneme} & \multirow{2}{*}[-0.5ex]{Approach} & \multicolumn{9}{c|}{Subject} & \multirow{2}{*}[-0.5ex]{Average} \\ \cmidrule(lr){3-11}
 &  & 1 & 2 & 3 & 4 & 5 & 6 & 7 & 8 & 9 &  \\ \midrule
\multirow{4}{*}{Initial} & Single-modality & 9.53 & 9.88 & 10.04 & 6.67 & 10.61 & 8.10 & 5.62 & 11.03 & 11.38 & 9.21 \\
 & Mel & 10.24 & 10.78 & 12.38 & 7.79 & 11.42 & 9.13 & 7.78 & 12.64 & 13.28 & 10.61 \\
 & wav2vec2 & \textbf{10.92} & \textbf{12.38} & 12.82 & \textbf{8.78} & 12.38 & 8.82 & \textbf{8.16} & \textbf{14.71} & \textbf{15.98} & \textbf{11.66} \\
 & HuBERT & 10.78 & 12.26 & \textbf{12.99} & 8.71 & \textbf{12.70} & \textbf{9.23} & 8.10 & 14.35 & 15.67 & 11.64 \\ \midrule
\multirow{4}{*}{Initial\_8} & Single-modality & 17.86 & 18.06 & 19.46 & 14.32 & 19.39 & 18.30 & 17.78 & 20.87 & 21.33 & 18.60 \\
 & Mel & 18.87 & 19.31 & 20.56 & 14.82 & 19.77 & 18.72 & 19.03 & 22.41 & 22.68 & 19.57 \\
 & wav2vec2 & \textbf{20.74} & 20.90 & 21.74 & \textbf{15.79} & \textbf{20.78} & \textbf{19.24} & \textbf{19.86} & 23.35 & \textbf{24.97} & \textbf{20.82} \\
 & HuBERT & 19.82 & \textbf{21.07} & \textbf{21.80} & 15.70 & 20.46 & 18.91 & 19.49 & \textbf{23.75} & 24.82 & 20.65 \\ \midrule
\multirow{4}{*}{Final} & Single-modality & 29.54 & 28.89 & 30.96 & 28.25 & 33.30 & 28.47 & 30.59 & 32.28 & 35.68 & 30.89 \\
 & Mel & 29.60 & 28.86 & 31.43 & 28.61 & 34.26 & 29.57 & 31.48 & 33.82 & 35.33 & 31.44 \\
 & wav2vec2 & 29.82 & \textbf{30.04} & \textbf{32.39} & 29.56 & 33.52 & 29.44 & \textbf{31.49} & 34.52 & \textbf{37.26} & 32.00 \\
 & HuBERT & \textbf{29.97} & 29.77 & 31.93 & \textbf{29.68} & \textbf{34.38} & \textbf{29.89} & 30.78 & \textbf{34.67} & 37.26 & \textbf{32.04} \\ \midrule
\multirow{4}{*}{Tone} & Single-modality & 59.29 & 57.28 & 52.82 & 46.81 & 42.74 & 27.66 & 30.51 & 39.01 & 45.34 & 44.61 \\
 & Mel & 59.07 & 56.99 & 53.44 & 47.26 & 44.05 & 28.82 & 33.47 & 40.62 & 46.08 & 45.53 \\
 & wav2vec2 & \textbf{60.81} & \textbf{59.28} & \textbf{55.86} & \textbf{49.36} & \textbf{46.83} & \textbf{29.17} & \textbf{34.41} & 41.93 & \textbf{49.78} & \textbf{47.49} \\
 & HuBERT & 59.81 & 58.50 & 55.08 & 48.99 & 46.03 & 28.38 & 33.80 & \textbf{41.98} & 48.71 & 46.81 \\ \bottomrule
\end{tabular}}
\begin{tablenotes}
\footnotesize
\item \textasteriskcentered ~The results of Initial\_8 and Final are measured by BCA.
\end{tablenotes}
\end{table*}

\subsection{Data Augmentation}

Table~\ref{table:single aug} shows top-5 and top-1 accuracies (\%) of within-subject single-modality decoding, with and without data augmentation. The best result in each column of panel is highlighted in bold. All four data augmentation approaches were effective. Among them, Salt \& Pepper noise performed the best in both domains, improving the average top-5 accuracy by up to 4.22\%, corresponding to a 4.22/25.86=16.32\% relative improvement, and the top-1 accuracy by up to 1.54\%, corresponding to a 1.54/6.65=23.16\% relative improvement.

\begin{table*}[htbp]    \centering  \setlength{\tabcolsep}{0.47mm}
\caption{Top-5 and top-1 accuracies (\%) of within-subject single-modality decoding, with and without data augmentation.}    \label{table:single aug}
\renewcommand{\arraystretch}{1}
\scalebox{0.6}{
\begin{tabular}{@{}cc|cccccccccccccccccc|cc@{}}
\toprule
\multicolumn{2}{c|}{\multirow{3}{*}[-1.5ex]{Approach}} & \multicolumn{18}{c|}{Subject} & \multicolumn{2}{c}{\multirow{2}{*}[-0.5ex]{Average}} \\ \cmidrule(lr){3-20}
\multicolumn{2}{c|}{} & \multicolumn{2}{c}{1} & \multicolumn{2}{c}{2} & \multicolumn{2}{c}{3} & \multicolumn{2}{c}{4} & \multicolumn{2}{c}{5} & \multicolumn{2}{c}{6} & \multicolumn{2}{c}{7} & \multicolumn{2}{c}{8} & \multicolumn{2}{c|}{9} & \multicolumn{2}{c}{} \\ \cmidrule(l){3-22}
\multicolumn{2}{c|}{} & Top5 & Top1 & Top5 & Top1 & Top5 & Top1 & Top5 & Top1 & Top5 & Top1 & Top5 & Top1 & Top5 & Top1 & Top5 & Top1 & Top5 & Top1 & Top5 & Top1 \\ \midrule
\multicolumn{2}{c|}{w/o augmentation} & 29.92 & 7.38 & 29.92 & 8.03 & 27.61 & 7.32 & 19.83 & 4.56 & 28.21 & 7.08 & 18.84 & 4.24 & 17.40 & 4.06 & 28.40 & 8.29 & 32.66 & 8.91 & 25.86 & 6.65 \\ \midrule
\multicolumn{1}{c|}{\multirow{4}{*}{Time noise}} & Gaussian & 30.15 & 7.75 & 29.74 & 8.08 & 28.19 & 7.60 & 20.76 & 4.71 & 28.56 & 7.25 & 19.67 & 4.40 & 17.44 & 3.90 & 28.85 & 8.32 & 33.03 & 9.31 & 26.26 & 6.81 \\
\multicolumn{1}{c|}{} & Poisson & 33.53 & 8.86 & 34.53 & 10.00 & 30.45 & 8.34 & 21.58 & 4.97 & \textbf{33.34} & 8.74 & 20.17 & 4.56 & 19.11 & 4.58 & 30.62 & 9.09 & 34.11 & 9.24 & 28.60 & 7.60 \\
\multicolumn{1}{c|}{} & Pink & 30.60 & 7.92 & 30.65 & 8.25 & 28.86 & 7.87 & 20.75 & 4.92 & 29.09 & 7.52 & 19.59 & 4.22 & 18.67 & 4.19 & 29.49 & 8.97 & 33.97 & 9.11 & 26.85 & 7.00 \\
\multicolumn{1}{c|}{} & Salt \& Pepper & \textbf{34.08} & \textbf{9.16} & \textbf{36.56} & \textbf{10.76} & \textbf{32.61} & \textbf{9.71} & \textbf{23.65} & \textbf{5.73} & 33.28 & \textbf{8.85} & \textbf{21.28} & \textbf{4.72} & \textbf{20.58} & \textbf{4.91} & \textbf{32.68} & \textbf{9.66} & \textbf{36.03} & \textbf{10.21} & \textbf{30.08} & \textbf{8.19} \\ \midrule
\multicolumn{1}{c|}{\multirow{4}{*}{Frequency noise}} & Gaussian & 30.31 & 7.81 & 29.54 & 8.03 & 28.42 & 7.67 & 20.49 & 4.82 & 28.65 & 7.14 & 19.48 & 4.20 & 18.21 & 4.15 & 29.29 & 8.76 & 32.95 & 8.93 & 26.37 & 6.83 \\
\multicolumn{1}{c|}{} & Poisson & 30.24 & 7.46 & 29.33 & 7.74 & 28.13 & \textbf{7.69} & \textbf{21.13} & 4.85 & 28.30 & 7.08 & 19.66 & 4.15 & 17.81 & 4.05 & 29.47 & 8.78 & 32.58 & 8.66 & 26.29 & 6.72 \\
\multicolumn{1}{c|}{} & Pink & 30.66 & 7.36 & 29.77 & 8.31 & \textbf{28.56} & 7.60 & 20.51 & 4.77 & 28.39 & 7.14 & 19.90 & 4.31 & 17.77 & 3.95 & 28.72 & 8.36 & 32.52 & 8.85 & 26.31 & 6.74 \\
\multicolumn{1}{c|}{} & Salt \& Pepper & \textbf{35.24} & \textbf{9.28} & \textbf{33.01} & \textbf{8.97} & 25.35 & 6.75 & 20.50 & \textbf{4.86} & \textbf{35.17} & \textbf{9.33} & \textbf{22.05} & \textbf{5.02} & \textbf{21.88} & \textbf{5.28} & \textbf{30.81} & \textbf{8.81} & \textbf{36.22} & \textbf{9.62} & \textbf{28.91} & \textbf{7.55} \\ \bottomrule
\end{tabular}
}
\end{table*}

Table~\ref{table:multi aug} presents the MASD decoding accuracies when using Salt \& Pepper noise in data augmentation. The best average result in each feature combination is underlined, and the best performance in each column is highlighted in bold. Data augmentation further improved the MASD decoding accuracies. The average top-5 accuracy reached 35.80\%, and the average top-1 accuracy reached 10.69\%. Specifically, for subject 9, the best-performing subject, the top-5 accuracy reached 46.21\%, and the top-1 accuracy reached 15.32\%.

\begin{table*}[htbp]  \centering  \setlength{\tabcolsep}{0.5mm}
\caption{Top-5 and top-1 accuracies (\%) in within-subject decoding, using MASD and different data augmentation approaches.}     \label{table:multi aug}
\renewcommand{\arraystretch}{1}
\scalebox{0.57}{
\begin{tabular}{@{}cc|cccccccccccccccccc|cc@{}}
\toprule
\multicolumn{2}{c|}{\multirow{3}{*}[-1.5ex]{Approach}} & \multicolumn{18}{c|}{Subject} & \multicolumn{2}{c}{\multirow{2}{*}[-0.5ex]{Average}} \\ \cmidrule(lr){3-20}
\multicolumn{2}{c|}{} & \multicolumn{2}{c}{1} & \multicolumn{2}{c}{2} & \multicolumn{2}{c}{3} & \multicolumn{2}{c}{4} & \multicolumn{2}{c}{5} & \multicolumn{2}{c}{6} & \multicolumn{2}{c}{7} & \multicolumn{2}{c}{8} & \multicolumn{2}{c|}{9} & \multicolumn{2}{c}{} \\ \cmidrule(l){3-22}
\multicolumn{2}{c|}{} & Top5 & Top1 & Top5 & Top1 & Top5 & Top1 & Top5 & Top1 & Top5 & Top1 & Top5 & Top1 & Top5 & Top1 & Top5 & Top1 & Top5 & Top1 & Top5 & Top1 \\ \midrule
\multicolumn{1}{c|}{\multirow{3}{*}{fastText+Mel}} & w/o augmentation & 36.20 & 9.81 & 37.84 & 11.63 & 34.53 & 10.33 & 26.51 & 6.74 & 35.78 & 10.20 & 22.22 & 5.02 & 24.07 & 5.99 & 36.68 & 12.24 & 41.67 & 12.85 & 32.83 & 9.42 \\
\multicolumn{1}{c|}{} & Time noise & 37.45 & 10.23 & 40.22 & 12.13 & 35.72 & 10.67 & 27.28 & 6.79 & 37.16 & 10.57 & \textbf{23.43} & 5.37 & 25.10 & 6.46 & 37.35 & 12.58 & 42.93 & 12.53 & \underline{34.07} & \underline{9.70} \\
\multicolumn{1}{c|}{} & Frequency noise & 36.94 & 9.76 & 40.22 & 12.91 & 35.82 & 10.76 & 25.76 & 6.22 & 37.72 & 10.73 & 22.35 & 5.15 & 24.24 & 6.12 & 35.91 & 11.42 & 42.58 & 12.76 & 33.50 & 9.54 \\ \midrule
\multicolumn{1}{c|}{\multirow{3}{*}{fastText+wav2vec2}} & w/o augmentation & 36.66 & 9.91 & 39.50 & 11.58 & 34.51 & 10.38 & 26.86 & 6.63 & 35.97 & 10.08 & 22.68 & 5.18 & 24.40 & 6.04 & 38.50 & 12.33 & 43.76 & 14.02 & 33.65 & 9.57 \\
\multicolumn{1}{c|}{} & Time noise & 37.37 & 10.26 & 41.19 & 12.81 & 35.78 & 11.19 & 27.50 & 7.18 & 38.00 & 10.89 & 22.66 & 5.65 & 25.52 & 6.34 & 38.17 & 12.75 & 44.99 & 14.15 & 34.57 & 10.14 \\
\multicolumn{1}{c|}{} & Frequency noise & 37.62 & 10.38 & 44.26 & 14.08 & 37.81 & \textbf{11.95} & 27.71 & 7.09 & 38.54 & 11.37 & 23.00 & 5.39 & 25.86 & 6.44 & 38.54 & 13.35 & 46.07 & 15.19 & \underline{35.49} & \underline{10.58} \\ \midrule
\multicolumn{1}{c|}{\multirow{3}{*}{fastText+HuBERT}} & w/o augmentation & 36.67 & 9.88 & 39.22 & 12.41 & 35.29 & 10.71 & 26.72 & 7.04 & 36.09 & 10.17 & 22.11 & 5.26 & 24.33 & 6.17 & 38.15 & 12.53 & 44.17 & 13.69 & 33.64 & 9.76 \\
\multicolumn{1}{c|}{} & Time noise & \textbf{38.08} & \textbf{10.59} & 40.94 & 12.53 & 35.33 & 11.38 & 27.79 & 6.98 & 37.48 & 10.81 & 22.64 & 5.43 & 25.45 & \textbf{6.51} & 38.81 & 13.14 & 44.32 & 13.83 & 34.54 & 10.13 \\
\multicolumn{1}{c|}{} & Frequency noise & 37.07 & 10.24 & 43.73 & 14.09 & 37.65 & 11.90 & 27.49 & 6.78 & 38.74 & 11.60 & 22.92 & 5.42 & 25.46 & 6.41 & 39.71 & 13.22 & 45.47 & 14.67 & \underline{35.36} & \underline{10.48} \\ \midrule
\multicolumn{1}{c|}{\multirow{3}{*}{BERT+Mel}} & w/o augmentation & 36.21 & 9.70 & 37.46 & 11.16 & 33.76 & 10.09 & 26.57 & 6.21 & 34.50 & 9.65 & 21.87 & 5.30 & 23.78 & 5.94 & 36.46 & 12.20 & 42.21 & 13.03 & 32.53 & 9.25 \\
\multicolumn{1}{c|}{} & Time noise & 37.06 & 10.47 & 39.82 & 12.28 & 35.20 & 10.86 & 26.92 & 6.69 & 36.27 & 10.20 & 22.92 & 5.14 & 24.61 & 6.01 & 37.22 & 12.42 & 42.71 & 13.22 & 33.64 & 9.70 \\
\multicolumn{1}{c|}{} & Frequency noise & 37.51 & 10.38 & 43.31 & 14.24 & 37.18 & 11.76 & 27.53 & 6.81 & 37.77 & 11.25 & 22.17 & 5.45 & 24.78 & 5.97 & 38.98 & 12.99 & 44.36 & 13.75 & \underline{34.84} & \underline{10.29} \\ \midrule
\multicolumn{1}{c|}{\multirow{3}{*}{BERT+wav2vec2}} & w/o augmentation & 37.08 & 10.03 & 39.53 & 12.29 & 34.72 & 10.67 & 27.01 & 6.86 & 36.86 & 10.63 & 22.41 & 5.26 & 24.27 & 6.12 & 38.21 & 13.00 & 44.22 & 14.56 & 33.81 & 9.94 \\
\multicolumn{1}{c|}{} & Time noise & 37.52 & 10.32 & 41.24 & 12.95 & 35.69 & 11.11 & 26.67 & 6.73 & 37.75 & 11.18 & 22.32 & \textbf{5.65} & 25.11 & 6.48 & 38.38 & 13.19 & 44.92 & 14.44 & 34.40 & 10.23 \\
\multicolumn{1}{c|}{} & Frequency noise & 37.65 & 10.44 & \textbf{44.44} & \textbf{14.44} & 37.60 & 11.88 & 27.58 & \textbf{7.18} & \textbf{39.20} & 11.38 & 23.33 & 5.48 & \textbf{26.15} & 6.28 & \textbf{40.05} & \textbf{13.80} & \textbf{46.21} & \textbf{15.32} & \underline{\textbf{35.80}} & \underline{\textbf{10.69}} \\ \midrule
\multicolumn{1}{c|}{\multirow{3}{*}{BERT+HuBERT}} & w/o augmentation & 37.02 & 9.74 & 39.22 & 12.13 & 34.76 & 10.49 & 26.24 & 6.89 & 36.08 & 10.53 & 22.39 & 5.36 & 23.65 & 5.86 & 38.13 & 13.34 & 44.55 & 14.06 & 33.56 & 9.82 \\
\multicolumn{1}{c|}{} & Time noise & 37.52 & 10.50 & 40.80 & 12.74 & 35.71 & 10.81 & 26.97 & 6.81 & 37.37 & 11.15 & 22.48 & 5.58 & 25.01 & 6.34 & 38.62 & 13.31 & 44.49 & 14.29 & 34.33 & 10.17 \\
\multicolumn{1}{c|}{} & Frequency noise & 38.05 & 10.28 & 44.26 & 14.31 & \textbf{38.01} & 11.84 & \textbf{28.03} & 7.00 & 38.74 & \textbf{11.74} & 22.66 & 5.31 & 25.52 & 6.35 & 39.97 & 13.69 & 45.35 & 14.99 & \underline{35.62} & \underline{10.61} \\ \bottomrule
\end{tabular}
}
\end{table*}

\subsection{Ablation Study}

\begin{figure}[htbp]\centering
\includegraphics[width=.5\linewidth,clip]{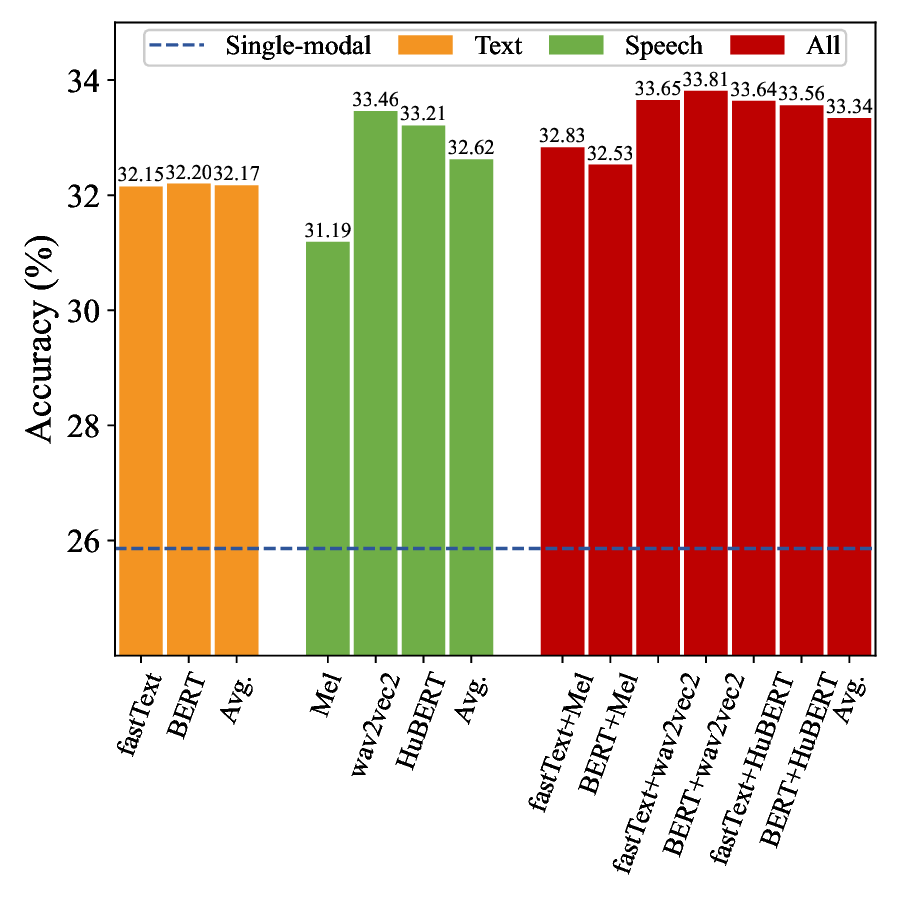}
\caption{Ablation study results.} \label{fig:ablation}
\end{figure}

To further investigate the effectiveness of using both text and synthetic speech for decoding, we employed only text or speech as assistance. Fig.~\ref{fig:ablation} shows the average results across the nine subjects:
\begin{enumerate}
\item Compared to the single-modal approach, using text or speech features as assistance significantly improved the decoding performance.
\item Compared to text-assisted decoding, the use of speech-assisted decoding with wav2vec2 and HuBERT showed considerable improvements.
\item Compared to using only a single assistive modality, using both text and synthetic speech assistance together further improved the decoding performance.
\end{enumerate}

\subsection{Parameter Sensitivity Analysis}

\begin{figure}[htbp]\centering
    \subfigure[]{\includegraphics[width=0.48\linewidth,clip]{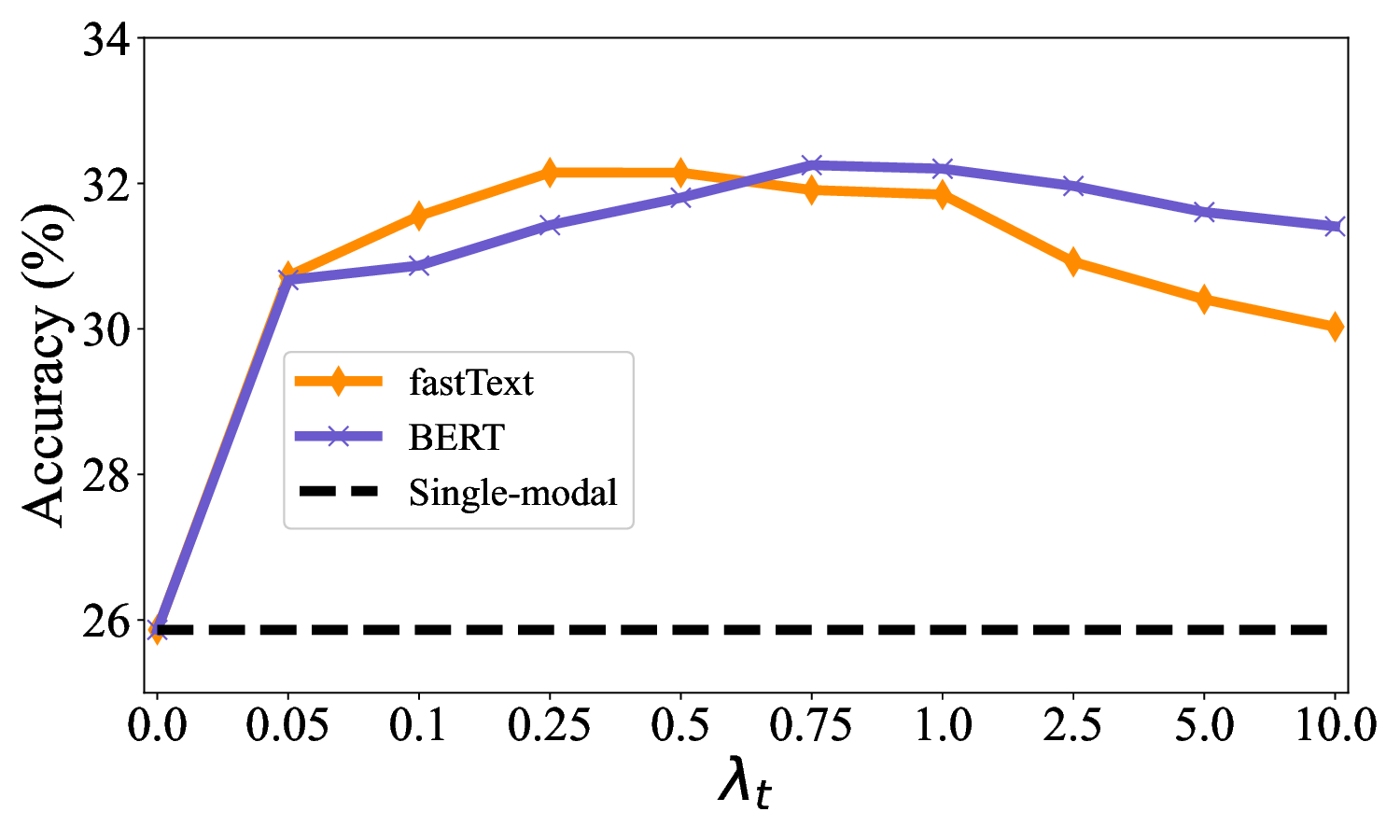}       \label{fig:lamb_t}}
    \hfill
    \subfigure[]{\includegraphics[width=0.48\linewidth,clip]{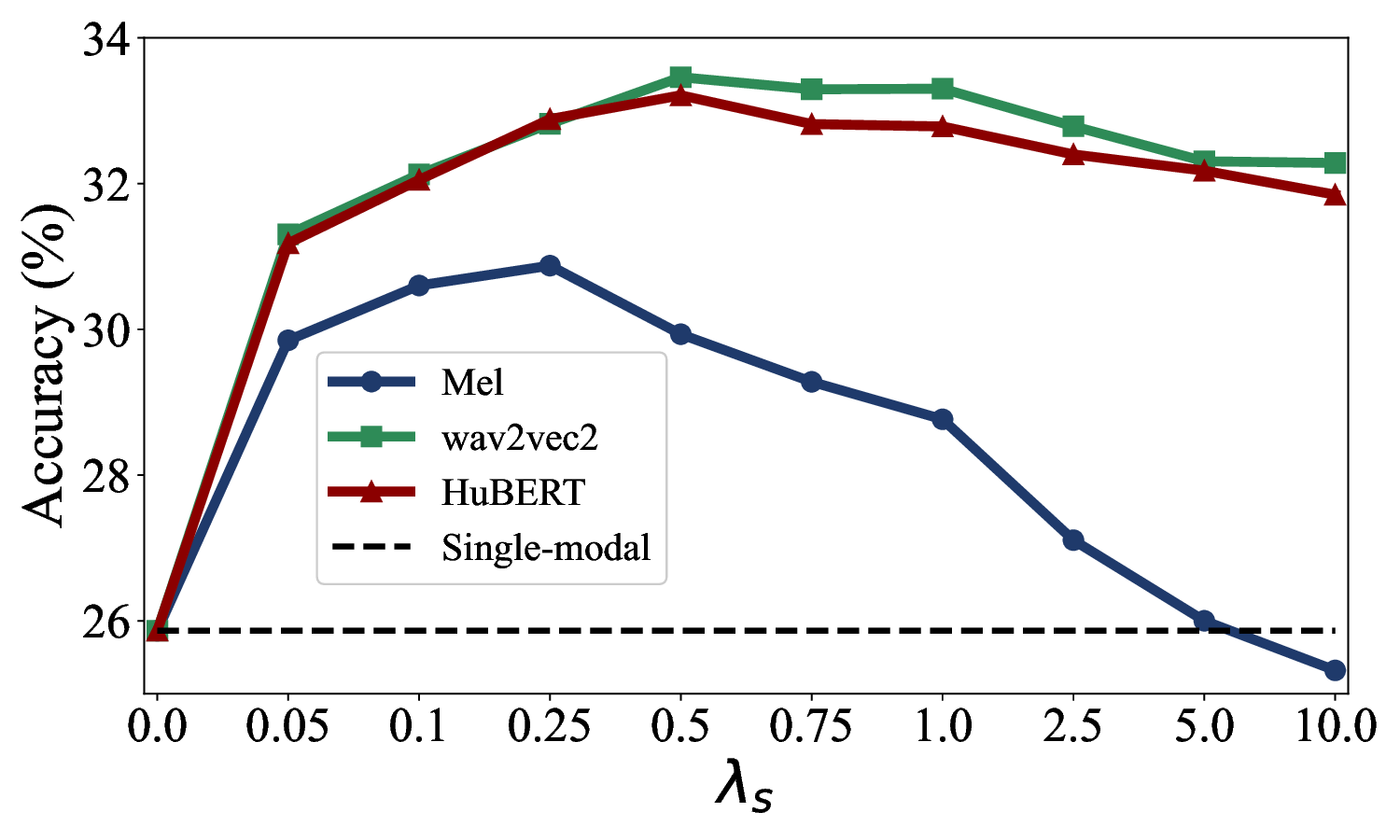}       \label{fig:lamb_s}}
    \caption{Top-5 decoding accuracies as the contrastive loss coefficients vary. \subref{fig:lamb_t} $\lambda_\text{t}$; \subref{fig:lamb_s} $\lambda_\text{s}$.} \label{fig:lamb}
\end{figure}

Fig.~\ref{fig:lamb} shows the top-5 accuracies when different contrastive loss weights $\lambda_\text{t}$ and $\lambda_\text{s}$ were used. To illustrate their effects more clearly, only a single assistive modality was used. It can be observed that different $\lambda_\text{t}$ and $\lambda_\text{s}$ consistently improved the decoding performance, when deep learning features were used.

\section{Conclusions} \label{sect:Conclusions}

Speech BCIs have the potential to directly reflect auditory perception and thoughts, offering a promising communication alternative for patients with aphasia. Chinese is one of the most widely spoken languages in the world, whereas there is very limited research on speech BCIs for Chinese language. This paper has acquired a dataset for non-invasive Chinese speech BCI study, which includes MEG signals collected from nine subjects based on a 48-word corpus. It also proposes a corresponding MASD algorithm for speech decoding, which utilizes both text and speech features to enhance the MEG feature extractor, significantly improving the decoding performance. Furthermore, we demonstrated that time or frequency domain data augmentation using various noise improved the decoding performance.

To our knowledge, this is the first study on modality-assisted decoding for non-invasive speech BCIs. Our future work will explore more challenging silent reading paradigms and more complex sentence decoding tasks.

\section*{Acknowledgement}

This research was supported by Wuhan Science and Technology Bureau under Grant 2024060702030150, the Taihu Lake Innovation Fund for Future Technology, HUST 2024-A-1, and the Fundamental Research Funds for the Central Universities 2023BR024.

\section*{References}

\bibliographystyle{IEEEtran}\bibliography{zhjiabib}

\end{document}